\documentstyle[preprint,aps]{revtex}
\begin{document}
\draft

\title{Power-Laws in Nonlinear Granular Chain under Gravity}
\author{Jongbae Hong, Jeong-Young Ji, and Heekyong Kim}
\address{Department of Physics Education, Seoul National University, Seoul 151-742, Korea}
\maketitle
\begin{abstract}
The signal generated by a  weak impulse  propagates in an oscillatory way and dispersively in 
a gravitationally compacted granular chain. For the power-law type contact force, 
we show analytically that the type of dispersion follows power-laws in depth. 
The power-law for grain displacement signal is given by $h^{-\frac{1}{4}(1-\frac{1}{p})}$ where 
$h$ and $p$ denote depth and the exponent of contact force, and the power-law for the grain 
velocity is $h^{-\frac{1}{4}(\frac{1}{3}+\frac{1}{p})}$. 
Other depth-dependent power-laws for oscillation frequency, wavelength, and period are given by 
combining above two and the phase velocity power-law $h^{\frac{1}{2}(1-\frac{1}{p})}$.  
We verify above power-laws by comparing with the data obtained by numerical simulations. 
\end{abstract}


\narrowtext 

Physics of granular  materials  attracts great   interest recently\cite{nagel}, since  these  materials are 
ubiquitous  around us and their properties are unique and also useful in many applications\cite{grain,powder}. 
The propagation of a sound or a weak elastic wave in a granular  medium is also  one of  interesting 
subjects related to the properties of  granular matter\cite{liu}.  A rather  simple system,  
the granular  chain with Hertzian contact\cite{landau},  has  been revived by finding a  soliton in 
transmitting elastic impulse. This soliton, existing in a highly nonlinear regime of a
horizontal Hertzian chain was first predicted  by Nesterenko\cite{nest} and 
its experimental verification was  performed by Lazaridi  and Nesterenko\cite{lazar} and  recently 
by Coste {\it et al}.\cite{coste}. Even though three-dimensional granular systems may not follow 
simple Hertzian contact force law due to geometrical effect\cite{goddard}, the one-dimensional granular chain 
with nonlinear contact force is still interesting. It may describe a fundamental feature of the
dynamics of nonlinear granular chain which appears in many areas of nature.
In addition, one-dimensional system is usually the starting point of studying higher dimensional systems. 

It is well-known that the velocity of an elastic impulse scales as $P^{1/6}$ or $h^{1/6}$ for the Hertzian 
chain\cite{goddard}, where $P$ is the pressure, linearly proportional to the depth $h$ for
vertical chain. Sinkovits and Sen\cite{sinko1} extended this to arbitrary nonlinear contact force of 
power-law type $F\propto \delta^p$, where $\delta$ denotes overlapped distance between adjacent 
grains. They showed that the signal velocity $v_{ph}$ scales as $h^{(1-\frac{1}{p})\frac{1}{2}}$ for 
$p\geq 1$ at large $h$. This has been simply obtained by considering the well-known relation 
$v_{ph}\propto\sqrt{\mu}$, where $\mu$ is the elastic constant which is given by $\mu\sim h^{1-p}$ 
for the above power-law type contact force. As far as we know, however, no power-law dependences on depth 
of the signal characteristics, such as oscillation frequency, period, and wavelength have been found 
in the gravitationally compacted chain.  

In this work, we study the propagation of acoustic or weak impulses in the  gravitationally 
compacted granular chain. We derive analytically the power-law behaviors of signal characteristics 
which depend on depth or time. We treat here a rather weak impulse which makes grain motion oscillatory 
and can be treated analytically even though it contains nonlinearity. The other extreme which is a
highly nonlinear regime has been studied by  Nesterenko\cite{nest}. 
Initial impulse may be used as a parameter which controls the solitariness of signal. 
The power-law behaviors for a wide range of impulse will be discussed in a separate
work\cite{heekyung}. 

We would like to obtain  analytically the exponents of various power-laws, such as  grain displacement, grain 
velocity, and oscillation period, frequency, and wavelength. 
We  first solve  the equation of motion of a grain displacement under gravity in the small oscillation or 
weak impulse regime in which the equation of motion under gravity can be mapped into 
the equation for the horizontal linear chain with varying force constant at each contact. 
The normal mode solution of the equation of motion can be obtained analytically in the continuum or 
long wavelength limit. The asymptotic behavior of the normal mode gives rise to the correct power-law 
behavior in depth, since the equation of motion has been changed into a linear form. 
Once we get the information on the grain velocity, all sorts of power-laws mentioned above can be obtained.
Since the equation of motion for grain velocity is not linear, the normal mode solution may not work to 
obtain power-law in depth. Therefore, we construct fully nonlinear forms describing displacement 
and velocity signal and obtain their depth-dependence behaviors. Our solution is quite general and gives 
rise to generic power-laws for arbitrary exponent $p$ of the contact force in the oscillating regime.

The equation of motion of $n$-th grain at $z_n$ is given by
\begin{eqnarray}
m\ddot{z}_n&=&\eta[\{\Delta_0-(z_n-z_{n-1})\}^{p}-\{\Delta_0-(z_{n+1}-z_{n})\}^{p}]
\nonumber \\ &+&mg,
\end{eqnarray}                                           
where $z_n$ is the distance from the top of chain to the center of the $n$-th spherical grain,
$m$ is the mass of grain, $\Delta_0$ is the distance between adjacent centers of the spherical grain,
and $\eta$ is the elastic constant of grain. 
Therefore, the overlap between adjacent grains at $n$th contact is $\delta_{n}=\Delta_0-(z_{n+1}-z_{n})$.
It is usually impossible to solve general nonlinear problems in an analytical way. Therefore we may not solve
the nonlinear differential equation of Eq. (1) exactly. But we may treat it analytically in a small 
oscillation regime which can be achieved by applying a weak impulse. 

For this purpose, we introduce a new variable 
\begin{equation}
\psi_n=z_n-n\Delta_0+\sum_{l=1}^n\left(\frac{mgl}{\eta}\right)^{1/p},
\label{vari}
\end{equation}                                                       
where the last term is the sum of overlaps up to the $n$th grain and set $z_{0}=\psi_0=0$. 
This change of variable makes Eq. (1) into an equation for the linearized horizontal chain 
with varying force constant at each contact, i.e.,
\begin{equation}
m\frac{\partial^2}{\partial t^2}\psi_n=-\mu_n(\psi_n-\psi_{n-1})+\mu_{n+1}(\psi_{n+1}-\psi_{n})
\label{linear}
\end{equation}
where $\mu_n=\mu_1n^{1-\frac{1}{p}}$ is the force constant of the $n$th contact and 
$\mu_1=mpg\left(\frac{\eta}{mg}\right)^{1/p}$ is the force constant of the first contact. 
Use of the condition of small oscillation 
\begin{equation}
|\psi_n-\psi_{n-1}|\ll\left(\frac{mgn}{\eta}\right)^{1/p}
\end{equation}
has been made to obtain Eq. (\ref{linear}) approximately. 
The expression of Eq. (\ref{linear}) in the continuum limit, i.e. 
the lattice constant $a=\delta h\rightarrow 0$, is given by
\begin{equation}
\frac{\rho}{\tau_1}\frac{\partial^2}{\partial t^2}\psi(h,t)=a^{\frac{1}{p}-1}\frac{\partial}{\partial h}
\left[\mu(h)\frac{\partial}{\partial h}\psi(h,t)\right]
\label{continuum}
\end{equation}
where $\mu(h)=h^{1-\frac{1}{p}}$ denotes the depth-dependence of force constant, 
and $\rho=m/a$ and $\tau_1=\mu_1a$ are the linear density and the tension of a chain at the first contact, 
respectively. We set $c_1=\sqrt{\tau_1/\rho}$ which is the well-known speed of wave in the string of 
tension $\tau_1$ and line density $\rho$.

We now choose $\psi_\zeta(h,t)=u_\zeta(h)e^{-ic_1a^{\frac{1}{2p}-\frac{1}{2}}\zeta t}$ as a normal mode 
solution. Then the depth-dependent function $u_\zeta(h)$ satisfies 
\begin{equation}
\frac{d^2}{dh^2}u_\zeta(h)+\frac{1-\frac{1}{p}}{h}\frac{d}{dh}u_\zeta(h)+
\frac{\zeta^2}{h^{1-\frac{1}{p}}}u_\zeta(h)=0,
\label{position}
\end{equation}                                                                 
which is a type of Bessel's differential equations\cite{abramo}.
If we consider a solution propagating to the positive $h$-direction, the solution of Eq. (\ref{position})
is given by the Hankel function\cite{abramo} 
\begin{equation}
u_\zeta(h)=h^\xi H_\nu^{(1)}(\theta h^\gamma),
\label{sol}
\end{equation}
where 
$\xi =\frac{1}{2p}, \, \gamma=\frac{1}{2}+\xi=\frac{1}{2}\left(1+\frac{1}{p}\right), 
\, \theta=\frac{\zeta}{\gamma}, \\ \nu=\frac{\xi}{\gamma}=\frac{1}{1+p}$.

The asymptotic form of Eq. (\ref{sol}) at large $h$ for a fixed $\nu$ is 
\begin{equation}
u_\zeta(h)\approx \sqrt{\frac{2}{\pi\theta}}h^{\xi-\frac{\gamma}{2}}
e^{i[\theta h^\gamma-\frac{\pi}{2}\nu-\frac{\pi}{4}]}
\end{equation}                                          
and the displacement function becomes 
\begin{equation}
\psi_\zeta(h,t)\sim h^{\xi-\frac{\gamma}{2}}e^{i\left[\frac{\zeta}{\gamma}h^\gamma-c_1
a^{\frac{1}{2p}-\frac{1}{2}}\zeta t\right]}
\label{asymp}
\end{equation}                                          

The amplitude of displacement signal is given by the envelope 
function of the asymptotic solution, which scales as 
\begin{equation}
A(h)\sim h^{\xi-\frac{\gamma}{2}}=h^{-\frac{1}{4}\left(1-\frac{1}{p}\right)}
\label{ampli}
\end{equation}          
The phase velocity $v_{ph}$ is obtained by setting the phase of Eq. (\ref{asymp}) constant,
i.e., $\frac{\zeta}{\gamma}h^\gamma-c_1a^{\frac{1}{2p}-\frac{1}{2}}\zeta t=$constant. We obtain
\begin{equation}
v_{ph}=\frac{dh}{dt}=c_1a^{\frac{1}{2}\left(\frac{1}{p}-1\right)}h^{\frac{1}{2}
\left(1-\frac{1}{p}\right)}
\label{phase}
\end{equation}
The group velocity scales as that of phase velocity and the dispersion relation is 
linear but depth-dependent in this case.

Since Eq. ({\ref{continuum}) is a linear differential equation, the envelope function of a normal mode 
solution in Eq. ({\ref{asymp}) can describe the depth-dependence of the displacement signal 
which may be given by the linear combination of all normal modes of different frequency. 
Therefore, one can expect that the depth-dependent behavior
predicted by Eq. ({\ref{ampli}) may agree well with the data given by numerical simulation which 
will be shown in what follows. 

The normal mode solution of Eq. ({\ref{asymp}) alone, however, cannot give appropriate predictions
on the changes in frequency, wavelength, and period of the signal as it propagates down. 
The information on the grain velocity may give the characteristics of signal 
dispersion by combining it with the displacement and velocity of signal.
To obtain the depth-dependence of grain velocity we write Eq. (\ref{continuum}) as 
\begin{equation}
\frac{\partial}{\partial t}v(h,t)=\frac{\partial}{\partial h}
\left[\mu(h)\frac{\partial}{\partial h}\psi(h,t)\right]
\label{velocity}
\end{equation}
We set the constant factor of  Eq. (\ref{continuum}) unity for our convenience, 
since it has nothing to do with depth-dependence behavior. 
To draw the depth-dependent behaviors of grain velocity, frequency, etc. 
we set the frequency $\omega(h)\sim h^\alpha$, the displacement function 
\begin{equation}
\psi(h,t)\sim h^{-\frac{1}{4}\left(1-\frac{1}{p}\right)}e^{ik(h)h-i\omega(h)t},
\label{dis} 
\end{equation}
and the grain velocity  
\begin{equation}
v(h,t)\sim h^{\beta}e^{ik(h)h-i\omega(h)t+\phi}, 
\label{vel}
\end{equation}
where $\alpha$ and $\beta$ will be determined and $\phi$ is the phase difference between displacement 
and velocity signal. The depth-dependent wavenumber $k(h)$ is given by $k(h)=v_{ph}(h)/\omega(h)$.

Since we have two unknowns $\alpha$ and $\beta$ to be determined, we need two independent equations for
these. One is given by Eq. (\ref{velocity}) and the other the relation $\omega(h)\sim v(h)/\psi(h)$.
Substituting Eqs. (\ref{dis}) and (\ref{vel}) into Eq. (\ref{velocity}) gives rise to
$\beta=\frac{1}{4}-\frac{5}{4p}-2\alpha$ and  the relation $\omega(h)\sim v(h)/\psi(h)$ yields 
$\alpha=\beta+\frac{1}{4}-\frac{1}{4p}$.
We obtain the power-law behaviors of grain velocity and frequency from these two equations as
follows:
\begin{eqnarray}
v(h)&\propto&h^{-\frac{1}{4}\left(\frac{1}{3}+\frac{1}{p}\right)} \\
\omega(h)&\propto&h^{\frac{1}{6}-\frac{1}{2p}}
\end{eqnarray}                   

The characteristic time of oscillation which is expressed by the period is given by the inverse of 
frequency or the ratio of displacement to grain velocity, i.e.,
\begin{equation}
T(h)= \frac{A(h)}{v(h)}\propto \omega(h)^{-1} \sim h^{-\frac{1}{6}+\frac{1}{2p}}
\label{time}
\end{equation}
The characteristic length of oscillation, on the other hand, which is expressed by wavelength is given by  
multiplying $T(h)$ by phase velocity, i.e., 
\begin{equation}
\lambda(h)=T(h)v_{ph}(h)\sim h^{1/3}
\label{time}
\end{equation}                                       
Interestingly enough, characteristic length of oscillation does not depend on contact force within 
the linear approximation of Eq. (3).
    
We now compare above results with molecular dynamics simulations performed for Eq. (1) for arbitrary $p$.
To perform numerical simulation for Eq. (1), we choose a vertical chain of $N=2\times10^3$ grains
and neglect plastic deformation. 
As a calculational tool, we use the third-order Gear predictor-corrector algorithm\cite{gear}.
We choose $10^{-5}$m, $2.36\times10^{-5}$kg, and $1.0102\times 10^{-3}$s as the units of distance, 
mass, and time, respectively. These units gives the gravitational acceleration $g=1$.  
We set the grain diameter 100, mass 1, and the elastic constant $\eta$ of Eq. (1) is given by 
$\eta=(1+p)b$, where $b$ depending on modulus is chosen as 5657 for this molecular dynamics 
simulation. The equilibrium condition
\begin{equation}
mgn=\eta\delta^{p}_{n}
\end{equation}                                      
has been used for the $(n+1)$th grain of a vertical chain. 
Even though a criterion\cite{coste,johnson} for initial impulse to 
neglect plastic deformation and viscoelastic dissipation in experimental situation, we do not 
care about that criterion for the numerical simulation. For the purpose of this work, however, 
we choose a rather weak initial impulse $v_i=0.1$ in our program units. There is a regime of 
initial impulse in which the signal follows the same power-laws. This will be shown in a separate 
paper\cite{heekyung}. 

Figure 1 shows the snap shots of amplitudes (a) and corresponding grain velocity signals (b) 
propagating down the vertical chain with Hertzian contact ($p=3/2$). The leading amplitude of displacement 
of each signal in Fig. 1(a) corresponds to the leading part of each velocity signal in Fig. 1(b). 

We focus on the leading amplitudes of displacement and velocity signal 
and plot them in log$_{10}$-log$_{10}$ scale in Fig. 2 which shows that both displacement 
and velocity peak decreases in power-laws of depth. 
The explicit expressions for the depth-dependent behaviors of leading amplitudes of displacement 
and velocity are given by $A_{max}(h)\propto h^{-0.0835\pm0.0003}$  
and $v_{max}(h)\propto h^{-0.2500\pm0.0001}$. 
We also obtain other depth-dependent power-laws showing dispersiveness of the signal. 
They are the elasped time to reach to maximum amplitude, $T_{max}(h)$, which describes the period and 
the number of particles participating at the leading part of velocity signal, $N(h)$, which describes 
the wavelength. The power-law exponents of these quantities with error bounds are 
$T_{max}\sim h^{0.170\pm0.002}, \, \, N\sim h^{0.338\pm0.004}$. These values are in good agreement with 
theoretical predictions for the Hertzian $p=3/2$.

We obtain peak values of displacement and grain velocity signal for other values of $p$ and 
plot them in Fig. 3. One can see a very nice fit to the theoretical curves up to $p=2$. For large 
values of $p$, the deviation from theory occurs especially in grain velocity. This is understandable 
because nonlinearity becomes stronger as $p$ increases and grain velocity contains more 
nonlinearity than displacement.  

In conclusion, the propagating feature in vertical chain is dispersive due to gravity 
even though total energy and momentum are conserved. The effect of gravity induces 
the change in force constant at every contact. Therefore, the signal is no longer a soliton which is 
the propagating mode in horizontal chain\cite{nest,coste}.  We treat the problem analytically 
for arbitrary power-law type of nonlinear contact forces and obtain general features of dispersive 
phenomena for a weak impulse in a gravitationally compacted chain. The normal mode solution for 
displacement has been obtained in the small oscillation and continuum limit. This normal mode 
solution describes the depth-dependent power-law of displacement signal. 
We find various power-laws describing signal characteristics depending on depth.  
This dispersive property is obtained by constructing both displacement and 
grain velocity function appropriately. 
We also perform numerical simulations for various power-law type contact force
for a rather weak impulse $v_i=0.1$ and show that the results agree well with our theoretical predictions.

The general features of soliton damping due to gravity may be given by studying 
similar work for a wide range of impulse, which will be given in a separate paper\cite{heekyung}.  
The properties of signal propagation studied in this work are fundamentals of the dynamics of 
granular chain under gravity. This work may be extended to higher dimensions and to more practical models 
for applications.  

One of authors (J.H.) thanks Prof. S. Sen of State University of New York at Buffalo and Prof. M. H.
Lee of the University of Georgia for useful discussions on this work. The authors wish to acknowledge
the financial support (1998-015-000055) of the Korea Research Foundation made in the program year of 1998.

\newpage
{\bf Figure Captions}

\begin{description}
\item[Fig. 1] :  (a) Snap  shots of displacement of the propagating wave in  a gravitationally  compacted
granular chain with Hertzian contact force law. Initial impulse is $v_i=0.1$.
(b) Snap  shots of grain velocity corresponding to (a).
\item[Fig. 2] :  Log$_{10}$-log$_{10}$ plots of leading peaks of displacement and grain velocity
shown in Fig. 1. Slopes of the straight lines are $-0.0835$ (displacement) and $-0.250$ (velocity),
respectively.
\item[Fig. 3] : Comparison of theory and simulation for power-law exponents for various $p$.
Simulation data are obtained for the leading peaks of displacement and grain velocity
signal.
\end{description}


\begin{references} 
\bibitem{nagel} H. M. Jaeger and S. R. Nagel, Science {\bf 255},  1523 (1995);  
H. M. Jaeger, S. R. Nagel, and  R. P. Behringer, Rev. Mod. Phys. {\bf 68}, 1259 (1996).
 
\bibitem{grain} {\it Physics of Dry Granular Media}, edited by H. J. Herrmann, 
J.-P. Hovi, and S. Luding (Kluwer Academic Publishers, Dordrecht, 1998). 

\bibitem{powder} {\it Powders and Grains 97}, edited by R. P. Behringer and 
J. T. Jenkins (A. A. Balkema, Rotterdam, 1997). 
 

\bibitem{liu} M. Leibig, Phys. Rev. E {\bf 49}, 1647 (1994); S. Melin, Phys. Rev. E {\bf 49}, 
2353 (1994); C-h. Liu, Phys. Rev. B {\bf 50}, 782 (1994); 
C-h. Liu and S. R. Nagel, Phys. Rev. B {\bf 48}, 15646 (1993); Phys. Rev. Lett. {\bf 68}, 2301 (1992).

\bibitem{landau} L.  D. Landau  and E.  M. Lifshitz,  {\it  Theory of  Elasticity} (Pergamon, 
Oxford, 1970), p. 30;
H. Hertz, J. Math. {\bf 92}, 156 (1881). 
                                              
\bibitem{nest} V. F. Nesterenko, J. Appl. Mech. Tech. Phys. (USSR) {\bf 5}, 733 (1984); 
V. F. Nesterenko, J. Phys. IV {\bf 55}, C8-729 (1994).

\bibitem{lazar} A. N. Lazaridi and V. F. Nesterenko, J. Appl. Mech. Tech. Phys.  (USSR) {\bf 
26}, 
405 (1985). 

\bibitem{coste} C. Coste, E. Falcon, and S. Fauve, Phys. Rev. E {\bf  56}, 6104 (1997).

\bibitem{goddard} J. D. Goddard, Proc. R. Soc. London, Ser. A {\bf 430}, 105 (1990). 

\bibitem{sinko1} R. S. Sinkovits and S. Sen, Phys. Rev. Lett. {\bf 74}, 2686 (1995);
S. Sen and R. S. Sinkovits, Phys. Rev. E {\bf 54}, 6857 (1996).

\bibitem{heekyung} J. Hong, H. Kim, and J.-P. Hwang (to be published).

\bibitem{abramo} M. Abramowitz and I. A. Stegun,  {\it Handbook of Mathematical Functions}, 
 AMS 55 (National Bureau of Standards, 1972). 

\bibitem{gear} M. P. Allen  and  D. J. Tildesley, {\it  Computer  Simulation  of Liquids} 
(Clarendon, Oxford, 1987).  

\bibitem{johnson} K. L. Johnson, {\it  Contact Mechanics} (Cambridge University, Press, 
Cambridge, 1992). 

\end{references}
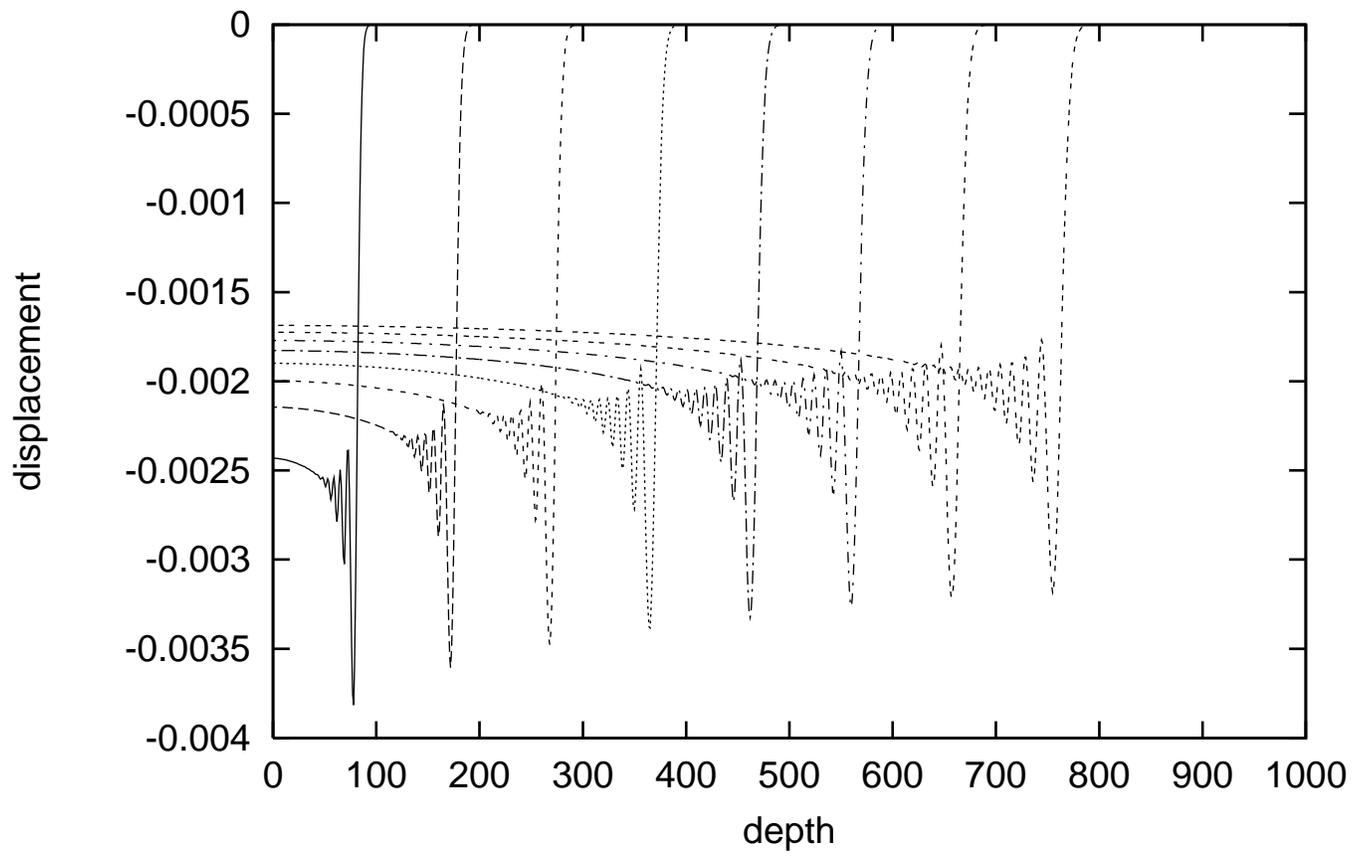
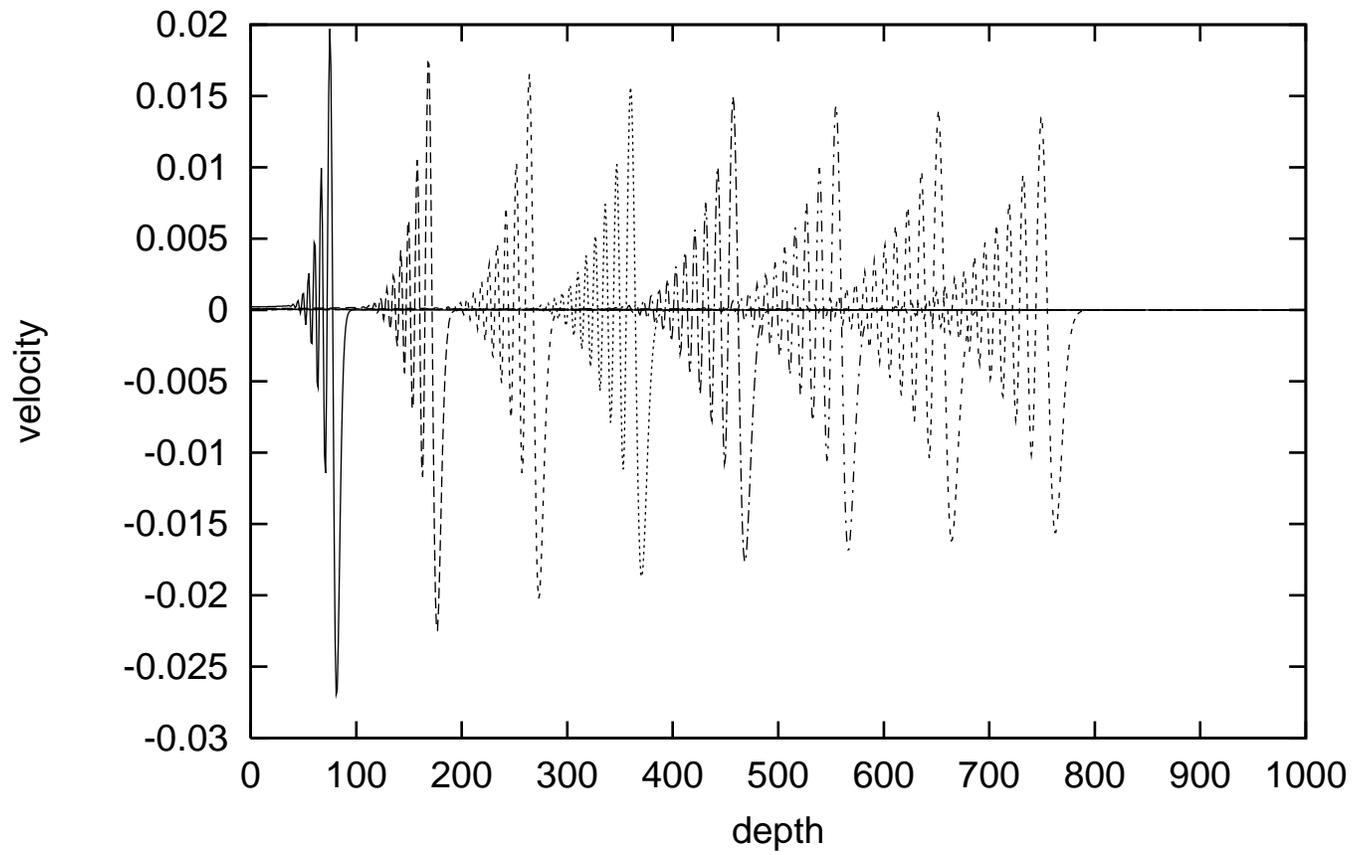
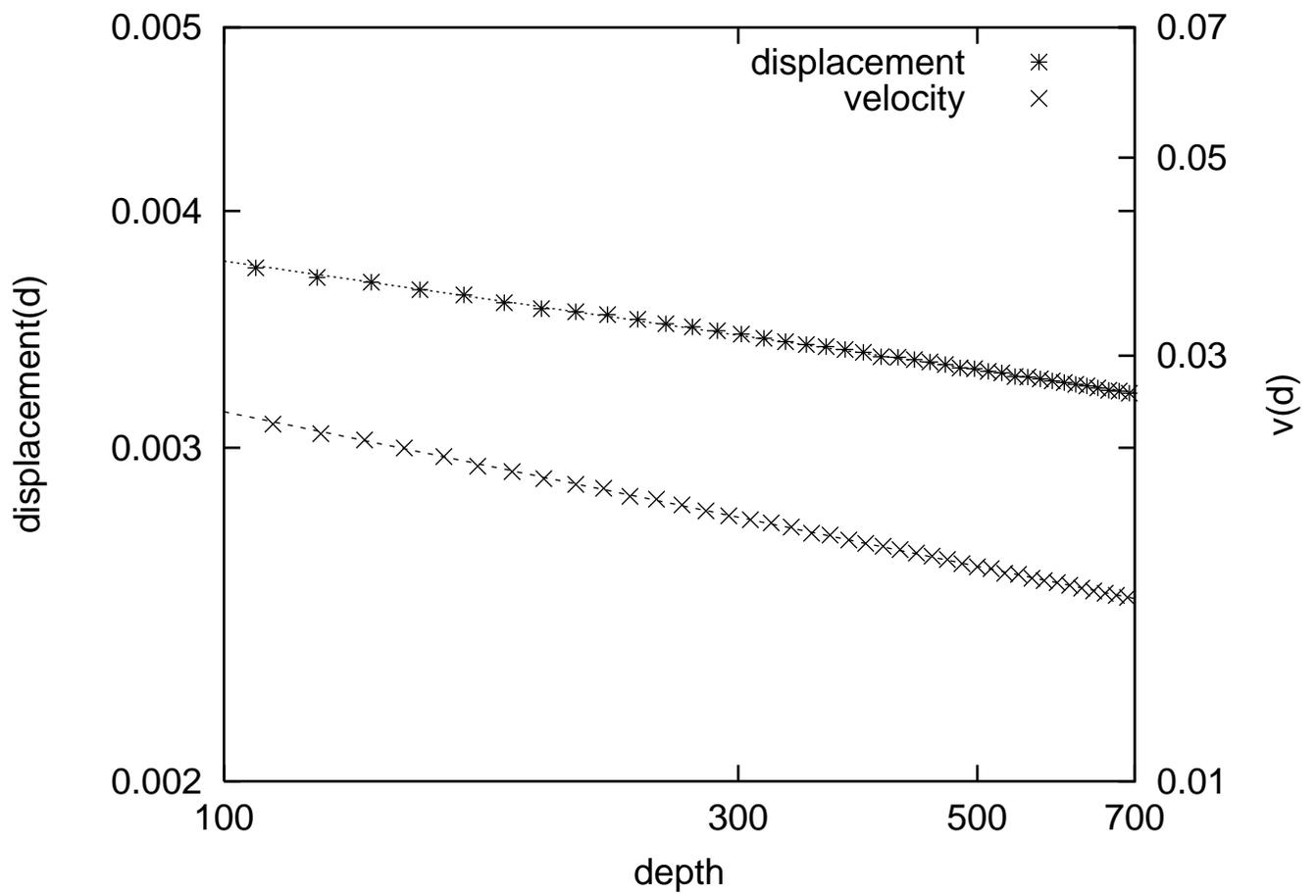
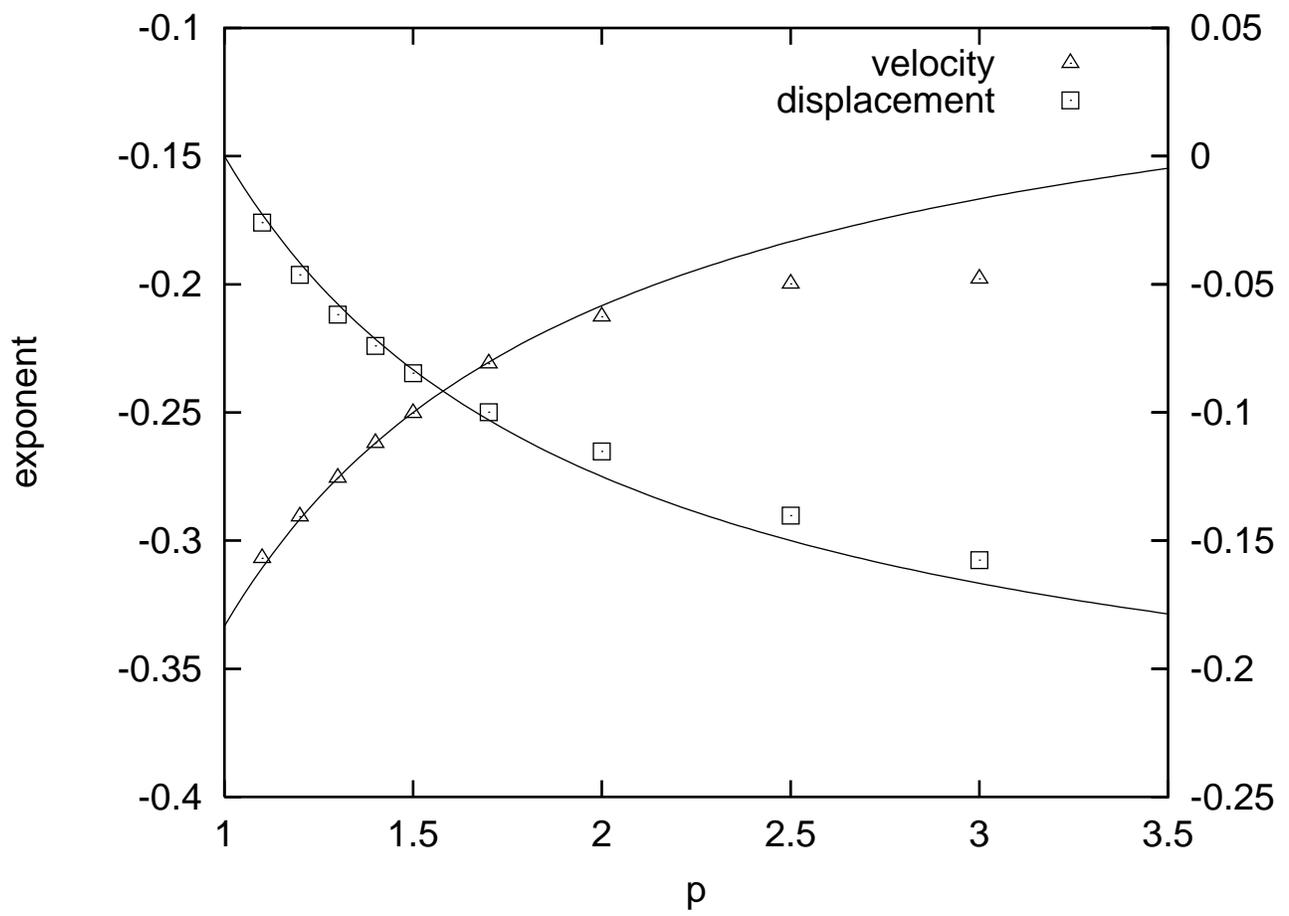
\end{document}